\documentstyle[12pt]{article}

\catcode`\@=11
\@addtoreset{equation}{section}
\@addtoreset{footnote}{section}

\def\section{\@startsection {section}{1}{\z@}{-3.5ex plus -1ex minus
 -.2ex}{2.3ex plus .2ex}{\large\bf}}
\def\subsection{\@startsection{subsection}{2}{\z@}{-3.25ex plus -1ex minus
 -.2ex}{1.5ex plus .2ex}{\bf}}
\def\subsect#1{\par\penalty1000{\noindent \bf #1}\par\penalty500}
\def\appendix{\par \setcounter{section}{0} \setcounter{subsection}{0}
 \def\thesection{\Alph{section}}\section}

\def\eqalign#1{\null\,\vcenter{\openup\jot\m@th
  \ialign{\strut\hfil$\displaystyle{##}$&$\displaystyle{{}##}$\hfil
      \crcr#1\crcr}}\,}
\def\eqalignno#1{\displ@y \tabskip\centering
  \halign to\displaywidth{\hfil$\@lign\displaystyle{##}$\tabskip\z@skip
    &$\@lign\displaystyle{{}##}$\hfil\tabskip\centering
    &\llap{$\@lign##$}\tabskip\z@skip\crcr
    #1\crcr}}

\catcode`\@=12

\def\eq#1{.~(\ref{#1})}
\def\beq{\begin{equation}}
\def\eeq{\end{equation}}
\def\beqar{\begin{eqnarray}}
\def\eeqar{\end{eqnarray}}

\def\nfrac#1#2{{\displaystyle{\vphantom1\smash{\lower.5ex\hbox{\small$#1$}}%
	\over\vphantom1\smash{\raise.25ex\hbox{\small$#2$}}}}}

\def\u#1{{}^{#1}}
\def\d#1{{}_{#1}}
\def\p#1{\mskip#1mu}
\def\n#1{\mskip-#1mu}
\def\stop{\p6.}
\def\comma{\p6,}
\def\semi{\p6;}
\def\and{\p6 {\rm and}}
\def\and{\p6 {\rm and}}
\def\Tr{{\rm Tr}}

\def\noj#1,#2,{{\bf #1} (19#2)\ }
\def\jou#1,#2,#3,{{\sl #1\/ }{\bf #2} (19#3)\ }
\def\ann#1,#2,{{\sl Ann.\ Physics\/ }{\bf #1} (19#2)\ }
\def\cmp#1,#2,{{\sl Comm.\ Math.\ Phys.\/ }{\bf #1} (19#2)\ }
\def\cq#1,#2,{{\sl Class.\ Quantum Grav.\/ }{\bf #1} (19#2)\ }
\def\cqg#1,#2,{{\sl Class.\ Quantum Grav.\/ }{\bf #1} (19#2)\ }
\def\ijmp#1,#2,{{\sl Int.\ J.\ Mod.\ Phys.\/ }{\bf A#1} (19#2)\ }
\def\jmp#1,#2,{{\sl J.\ Math.\ Phys.\/ }{\bf #1} (19#2)\ }
\def\grg#1,#2,{{\sl Gen.\ Rel.\ Grav.\/ }{\bf #1} (19#2)\ }
\def\mpl#1,#2,{{\sl Mod.\ Phys.\ Lett.\/ }{\bf A#1} (19#2)\ }
\def\nc#1,#2,{{\sl Nuovo Cim.\/ }{\bf #1} (19#2)\ }
\def\np#1,#2,{{\sl Nucl.\ Phys.\/ }{\bf B#1} (19#2)\ }
\def\pl#1,#2,{{\sl Phys.\ Lett.\/ }{\bf #1B} (19#2)\ }
\def\pla#1,#2,{{\sl Phys.\ Lett.\/ }{\bf #1A} (19#2)\ }
\def\pr#1,#2,{{\sl Phys.\ Rev.\/ }{\bf #1} (19#2)\ }
\def\prd#1,#2,{{\sl Phys.\ Rev.\/ }{\bf D#1} (19#2)\ }
\def\prl#1,#2,{{\sl Phys.\ Rev.\ Lett.\/ }{\bf #1} (19#2)\ }
\def\prp#1,#2,{{\sl Phys.\ Rept.\/ }{\bf #1C} (19#2)\ }
\def\ptp#1,#2,{{\sl Prog.\ Theor.\ Phys.\/ }{\bf #1} (19#2)\ }
\def\rmp#1,#2,{{\sl Rev.\ Mod.\ Phys.\/ }{\bf #1} (19#2)\ }
\def\yadfiz#1,#2,#3[#4,#5]{{\sl Yad.\ Fiz.\/ }{\bf #1} (19#2) #3%
\ [{\sl Sov.\ J.\ Nucl.\ Phys.\/ }{\bf #4} (19#2) #5]}
\def\zh#1,#2,#3[#4,#5]{{\sl Zh.\ Exp.\ Theor.\ Fiz.\/ }{\bf #1} (19#2) #3%
\ [{\sl Sov.\ Phys.\ JETP\/ }{\bf #4} (19#2) #5]}
\def\ie{\hbox{\it i.e.\/}}
\def\etc{{\it etc.\/}}

\def\ibid{{\it ibid.\/}}

\def\pa{\partial}
\def\pb{\bar \pa}

\def\to{\rightarrow}
\def\implies{\Rightarrow}

\def\D{{\cal D}}
\def\Vp{V^{++}}
\def\Vm{V^{--}}
\def\P{P^{(-3)}}
\def\Pmu{P^{(-3)\mu}}
\def\Lam{\Lambda^{(+4)}}
\def\chif{\chi^{(-4)}}
\def\chib{\bar\chi^{(+4)}}
\def\dxf{d^{\p1 4}\n2 x}
\def\dpf{\frac{d^{\p1 4}\n2 p}{(2\pi)^4}}
\def\dlf{\frac{d^{\p1 4}\n2 l}{(2\pi)^4}}
\def\du{d^{\p1 2}\n2u}
\def\duone{d^{\p1 2}\n2u_1}
\def\dutwo{d^{\p1 2}\n2u_2}
\def\eps{\varepsilon}
\def\sutlr{SU(2)_L{\times}SU(2)_R}
\def\sutl{SU(2)_L}
\def\sutr{SU(2)_R}
\def\sltr{SL(2,R)}
\def\sof{SO(4)}
\def\sut{SU(2)}
\def\ut{U(2)}
\def\uo{U(1)}
\def\uor{U(1)_R}

\begin{document}
\begin{titlepage}

\begin{center}
December 5, 1991
\hfill    TAUP-1928-91

\vskip 1 cm

{\large \bf
Harmonic Space, Self-Dual Yang Mills and the $N=2$ String.
}

\vskip 1 cm

\vskip 1.2cm
{
          Neil Marcus, Yaron Oz {\it and\/}
	  Shimon Yankielowicz\footnote{
Work supported in part by the US-Israel Binational Science Foundation,
and the Israel Academy of Science.\\E-Mail:
NEIL@HALO.TAU.AC.IL, YARONOZ@TAUNIVM.BITNET, H75@TAUNIVM.BITNET.}
}
\vskip 0.2cm

{\sl
School of Physics and Astronomy\\Raymond and Beverly Sackler Faculty
of Exact Sciences\\Tel-Aviv University\\Ramat Aviv, Tel-Aviv 69978, ISRAEL.
}

\end{center}

\vskip 1 cm

\begin{abstract}

The geometrical structure and the quantum properties of the recently
proposed harmonic space action describing self-dual Yang-Mills (SDYM) theory
are analyzed. The geometrical structure that is revealed is closely related
to the twistor construction of instanton solutions. The theory gets no
quantum corrections and, despite having SDYM as its classical equation of
motion, its S matrix is trivial. It is therefore {\it not\/} the theory of
the $N=2$ string.  We also discuss the 5-dimensional actions
that have been proposed for SDYM.

\end{abstract}
\newpage

\end{titlepage}

\section{Introduction}

\indent

Self-Dual Yang-Mills (SDYM) theory appears to be a basic ingredient of
various areas of research in Physics and Mathematics. The instanton
solutions \cite{BPST,Ward,inst} provide a non-perturbative field theory
information, and are powerful tool for classifying smooth and exotic
four manifolds \cite{Donaldson}.  SDYM in spacetimes with signature
$(2,2)$ has recently appeared as the effective theory of the $N = 2$
heterotic string \cite{OVh} (after it has been reduced to two or
three dimensions), as well as the $N=2$ open string \cite{prep}.  In the
$N=2$ theories, three-point tree-level S-matrix elements are nontrivial, but
do not describe scattering, and the four-point S-matrix elements vanish
\cite{OV,OVh,prep}.  If, as one would expect, this is true also for the
higher-point classical S-matrix elements, this would prove the
widely-held belief that SDYM (and self-dual gravity) has no classical
scattering \cite{belief}, and is a four-dimensional integrable
system with infinite number of degrees of freedom \cite{integ4}. It has
been conjectured that all two-dimensional integrable systems are
reductions of SDYM \cite{conject} and, indeed, many integrable
systems in two dimensions have been derived via such reductions
\cite{integ2}.  Thus, SDYM may be a unifying master system for these
theories.

Until now work on SDYM has been concentrated mainly on studying the
classical equations of motion.  It would clearly be interesting to be
able to also study the quantum behavior of the theory.  This is
needed in order to be able to compare SDYM to the $N=2$ string at the
quantum level, and also may shed light on the poorly understood notion of
quantum integrability. Evidently, in order to proceed further one needs an
action for the theory.  One such action is written in terms of scalars on a
K\"ahler manifold \cite{NS}.  This action has been shown to produce the
tree-level scattering amplitudes of the $N=2$ heterotic string (at least
up to the four-point function) \cite{OVh}, but does not successfully
describe the loop amplitudes.  It has been conjectured that this may be
because the action does not preserve enough of the geometric structure and
symmetries of SDYM.

Because of this, there has been much interest in a harmonic space
action recently proposed by Kalitzin and Sokatchev \cite{Action}. In
this paper we first study the geometrical structure of SDYM in harmonic
space, and elucidate the connection between the harmonic and twistor
formalisms.  We see that the harmonic-space action does indeed possess
many of the geometrical properties desired for a SDYM theory.  We then
look at the quantum properties of the action, and show that its S
matrix is trivial.  The theory therefore does not describe the $N=2$
string. In the conclusion we discuss why this may be the case, and also
discuss the two five-dimensional formalisms for SDYM proposed by Nair
and Schiff \cite{NS}.

\section{Actions and the geometrical structure of SDYM}

\indent
As a preliminary to understanding the geometrical structure of the
harmonic space action, we first briefly describe the twistor
interpretation of instantons, \ie{} gauge fields with self-dual field
strengths $F$ \cite{Ward,inst}.  This geometrical description will then
be explicitly carried out using spinor notation, in order to connect it
to the more algebraic harmonic approach, following which we reach the
action of ref~\cite{Action}.

\subsect{Complex coordinates and twistors}

Any 2-form $F$ on $R^4$ can be decomposed into self-dual and
anti-self-dual forms $F^+$ and $F^-$ as
\beq
F = F^+ + F^-  \comma   \label{623}
\eeq
where $ ^*F^{\pm}={\pm}F^{\pm}$, with $*$ being the Hodge operator.  Group
theoretically, this corresponds to the decomposition of the reducible
6-dimensional representation of $SO(4)$ into a $3_L$ and $3_R$ of
$\sutlr \simeq \sof$.  In the twistor approach, one needs to introduce
complex coordinates on $R^4\simeq C^2$.  This means choosing a complex
structure on $R^4$, so the ``Lorentz'' group is reduced from $\sof$ to
$\ut$.  On a complex manifold the exterior derivative ${\rm d} \to \pa + \pb$,
so the 2-form $F$ naturally decomposes into
\beq
F = F^{(2,0)} + F^{(1,1)} + F^{(0,2)} \comma \label{comp}
\eeq
where $(\alpha, \beta)$ refer to the degree of the form with respect
to $(\pa,\pb)$.  If $F$ is self-dual then
\beq
F^{(2,0)} = 0 = F^{(0,2)} \comma \label{11}
\eeq
but the converse is not necessarily true.  This is because $F^{(1,1)}$ is
a reducible 4-dimensional representation of $\ut$, and can be further
decomposed as
\beq
F^{(1,1)} = F^{(1,1)}_{(0)} + {\mu} \stop \label{decomp}
\eeq
Now $F^{(1,1)}_{(0)}$, which is irreducible,
corresponds to the self-dual piece $F^+$, so $\mu$, which corresponds to
the metric, must vanish for $F$ to be self dual.

There are now two approaches to using this result to get equations
implying SDYM.  The first is simply to require that $F$ be a $(1,1)$
form with $\mu=0$.  This can be achieved by noting that eq\eq{11}
implies that
\beq
\eqalign{
A & = g^{-1} \pa g \cr
{\bar A} & = g^\dagger \pb g^{\dagger-1} \stop
}
\eeq
If $g$ is taken to be hermitian, which can be done by an appropriate gauge
choice, then the equation $\mu=0$ becomes
\beq
\eta^{\bar\mu\nu} \pa_{\bar\mu} (g^{-2} \pa_\nu g^2) ~=~0 \comma \label{EofM}
\eeq
which is the equation for SDYM originally written by Yang \cite{Yang}.
If one defines $g = e^{\phi/2}$, eq\eq{EofM} becomes the equation of
motion of the $N=2$ heterotic string \cite{OVh}, and also of the open
$N=2$ string \cite{prep}.  This equation of motion can be derived from
an action for $\phi$ that consists of an infinite series of
terms \cite{OVh}, but can be rewritten more elegantly in
five dimensions as a ``K\"ahler-Chern-Simons'' theory \cite{NS}.  One
can also write a five-dimensional action directly in terms of $A$ that
implies both that $F$ is a $(1,1)$ form and that $\mu=0$ \cite{NS}.
We shall discuss these actions in the conclusion of the paper.

The second approach is that taken in twistor theory.
Both the twistor and harmonic-space formalisms are naturally
written in Euclidean spacetimes with signature $(4,0)$, and the Wick
rotation to a $(2,2)$ signature\footnote{As is well known, one can not
have SDYM in Minkowski spacetime with signature $(3,1)$, unless one
relaxes the hermiticity of the vector potential.} is not
straightforward.  We shall therefore be in Euclidean space from now
on.  The twistor construction of instantons uses the fact that,
instead of explicitly demanding that $\mu$ vanishes, one can rather
require that $F$ be of type $(1,1)$ for all the complex structures
(with a fixed orientation) that one can introduce on $R^4$.  This is
true because changing the complex structure mixes $F^{(2,0)}$,
$F^{(0,2)}$ and $\mu$ so, if $F^{(2,0)}$ and $F^{(0,2)}$ always vanish,
only  $F^{(1,1)}_{(0)}$ can remain and $F$ must be self-dual.  Since
the set of complex structures on $R^4$ is $\sof/\ut \simeq S^2$, one
can find instantons by looking for $(1,1)$ forms on $R^4\times S^2$,
and projecting them back to $R^4$.  Thus one has the following theorem
(compactifying $R^4$ to $S^4$): {\em A 2-form $F$ on $S^4$ is self-dual
iff its lift to $CP^3\;\simeq\;S^4\times S^2$ is of type $(1,1)$}
\cite{Atiyah}.

At this stage one would like to write an action to implement these
ideas.  However, this is quite nontrivial.  One can find an action
whose equation of motion leads to $F$ being $(1,1)$ on a complex 3
manifold, but only if it is a Calabi-Yau manifold \cite{Witten}.  This
is not the case for the projective twistor space ($CP^3$), since the
first Chern class of the $S^2$ piece is not zero.  The action of
\cite{Witten} therefore contains a singularity on $CP^3$.  As we shall see
in the following, the harmonic-space action proposed in \cite{Action}
does result in $F$ being a $(1,1)$ form on $CP^3$, where each different
complex structure on $R^4$ is parameterized by a point in the fiber.
However, it does this in a rather indirect fashion, and the action does
not contain singularities.

\subsect{Spinor notation and twistors}

In order to clarify the content of the previous section, and to
continue to harmonic space, it is necessary to use spinor notation.
The four coordinates of spacetime are thus written as $x^{\mu i}$, with $\mu$
and $i$ indices of $\sutl$ and $\sutr$ respectively, and satisfy the
reality condition\footnote{Our conventions are $\eps_{12} = 1$,
${\eps}^{12} = 1$, $x_i  = \eps_{i j} x^{j}$, $x^j  =
\eps^{i j} x_i$.}
\beq
x^{\mu i}  = \eps^{i j} \eps^{\mu \nu} ( x^{\nu j} )^* \stop \label{real}
\eeq
Similarly, the covariant derivative is $D_{{\mu}i} \equiv {\pa}_{{\mu}i} +
A_{{\mu}i}$.  The Yang-Mills field strength now becomes
\beq
F_{{\mu}i,{\nu}j} \, \equiv \, [D_{{\mu}i},D_{{\nu}j}]
	\, \equiv \, {\eps}_{i j}F_{(\mu\nu)}^+ + {\eps}_{\mu\nu} F_{(i j)}^-
	\comma \label{fdef}
\eeq
where the last identity uses the fact that an antisymmetric 2 tensor
of $\sut$ must be proportional to the $\eps$ symbol.
Since $F_{(\mu\nu)}^+$ and $F_{(i j)}^-$ clearly transform as the $(3,1)$ and
$(1,3)$ representations of $\sutlr$, respectively, they are the
self-dual and anti-self-dual pieces of the curvature.  We thus see
explicitly the decomposition of eq\eq{623}, and see that the
self-duality equation is simply
\beq
F_{(i j)}^- = 0 \stop \label{sd}
\eeq

To see the meaning of this on $C^2$, we now introduce complex coordinates:
\beq
\eqalign{
z^\mu &\equiv x^{\mu 1} \cr
\bar{z}_\mu &= (z^\mu)^* = x_\mu\u2 = \eps_{\mu\nu}
x^{\nu 2} \stop \label{compdef}
}
\eeq
Since this definition does not break $\sutl$, we see that the complex Lorentz
group $\ut$ is $\sutl\times\uo_R$, where $\uo_R$, the diagonal subgroup of
$\sutr$, multiplies $z^\mu$ by a phase.  Using the definition
of $F$ in eq\eq{fdef}, we see that
\beq
\eqalign{
   F_{z^\mu,z^\nu}^{(2,0)} &~=~ \eps_{\mu\nu} \, F_{11}^- \cr
   F_{\bar{z}_\mu,\bar{z}_\nu}^{(0,2)} &~=~ -\eps^{\mu\nu} F_{22}^- \comma
}
\eeq
and
\beq
F_{12}^- ~=~ - \nfrac12\, F_{z^\mu,\bar{z}_\mu}^{(1,1)} \stop
\eeq
This last term is (proportional to) $\mu$
so, as promised, we see that self-duality
is equivalent to $F$ being a $(1,1)$ form with $\mu=0$.  Also, since
changes of complex structure mix $F_{11}^- $, $F_{12}^- $ and $F_{22}^-$,
one sees explicitly that if $F$ is a $(1,1)$ form for all complex
structures, it is self-dual.

For the twistor approach, we now want to work on a fiber bundle with
base space $R^4$ and all complex structures (with fixed orientation) on
$R^4$ as the fiber.  The
fiber is $\sutr/\uor \,\simeq\, S^2$.  This can be seen by considering
the space as the different ways of defining $z^\mu$, modulo the complex
Lorentz group.  Then the space is covered by rotating
$z^\mu$ by an $\sutr$ matrix $u^i\d{j}$ in the definition eq\eq{compdef}:
\beq
\eqalign{
z^\mu &\to x^{\mu j} \, u^1\d{j} \cr
      &= z^\mu \, u^1\d{1} - \eps^{\mu\nu} \, \bar{z}_\nu \, u^1\d{2} \comma
	\label{crot}
}
\eeq
modulo the $\uor$ of multiplication of $z^\mu$ by a phase.  To work in
twistor space, one now needs coordinates on the projective twistor space.
Unfortunately, by choosing an explicit parameterization of $S^2$ one
obscures the symmetries of the $S^2$ making it harder to understand the
resulting physics.

\subsect{Harmonic space}

In the harmonic approach\footnote{The harmonic space construction
presented here is that of ref~\cite{harmsdym}.  Only the geometric
interpretation, and the relation to twistor theory are original work.
For more details on harmonic theory, see ref~\cite{harmsdym} and
references therein.}
$S^2$ is instead represented as $\sut/\uo$, where a new $\uo$ is
introduced to achieve the coseting \cite{harm}.  The coordinates of
harmonic space are the four-dimensional coordinates $x^{\mu i}$, and
the harmonic coordinates $u^{{\pm}i}$, defined as
\beq
\left(\begin{array}{c} u^{+i}\\ u^{-i}\end{array}\right)\in \sut
\stop \label{udef}
\eeq
There are only three independent $u^{\pm i}$, parameterizing
$\sut$, since they satisfy the $\sut$ orthogonality condition
\beq
u^{+ i} u^-\d{i} \,\equiv\, \eps_{i j} \,u^{+ i} u^{- j} = 1 \stop
\label{orthog}
\eeq
Note that, unlike the usual approach to harmonic space, it is now
clear that the $\sut$ of the harmonic space is the same as the $\sutr$ of
spacetime, since the harmonic coordinates are
designed to give the space of complex structures of spacetime.
The $+$ and $-$ signs of the $u^{\pm i}$'s indicate their
transformations under the new $\uo$ transformation, which acts on the
$u$ matrix in eq\eq{udef} by left-multiplication.

The harmonic space is reduced to the fiber bundle $R^4 \times S^2$, by
imposing the condition that all fields in the space have a fixed $\uo$
charge.  A function $f^{(q)}(u)$ with $\uo$ charge $q$ can then by
defined by its harmonic expansion:
\beq
f^{(q)}(u)=\sum_{n=0}^{\infty}f^{(i_1 \ldots i_{n+q}j_1 \ldots j_n)}
\, u^+_{i_1} \ldots u^+_{i_{n+q}}u^-_{j_1} \ldots u^-_{j_n}
\stop \label{expan}
\eeq
Here the $f$'s are symmetric, since any antisymmetric
piece can be reduced using eq\eq{orthog}.  They are therefore
irreducible $\sut$ tensors.  Harmonic space therefore describes $S^2$ and
general functions on it without using an explicit parametrization of the
sphere.

One now wants to define harmonic-space integration and differentiation.
The only integration rule which is $\sut$ invariant is
\beq
\int \du \, 1  =1 \qquad \qquad \int \du \, u^+_{(i_1}\ldots
u^+_{i_m}u^-_{j_1}\ldots u^-_{j_n)}=0 \qquad (m+n\neq 0) \stop
\label{hint}
\eeq
Differentiation on $S^2$ is given in terms of the three
Lie-derivatives $D^{++}$, $D^{--}$ and $D^0$.  They are defined as
\beq
\eqalign{
	D^{++} & \equiv u^{+i}\frac{\pa}{{\pa}u^{-i}} \cr
	D^{--} & \equiv u^{-i}\frac{\pa}{{\pa}u^{+i}} \cr
	D^{0 \hphantom{M} }
       	& \equiv u^{+i}\frac{\pa}{{\pa}u^{+i}} -
		       u^{-i}\frac{\pa}{{\pa}u^{-i}} \stop \label{lie}
}
\eeq
Note that $D^0$ is the generator of the new $\uo$.
Finally, derivatives in the spacetime directions are lifted into the
fiber bundle $R^4{\times}S^2$ by defining
\beq
\eqalign{
	\pa_\mu^+ \, & \equiv \, u^{+ i} \pa_{\mu i} \cr
	\pa_\mu^- \, & \equiv \, u^{- i} \pa_{\mu i} \stop
}
\eeq
{\it The crucial feature of harmonic space is that $\pa_\mu^+$ and $\pa_\mu^-$
are Lorentz-covariant descriptions of $\pa_{z^\mu}$ and
$\pa_{\bar{z}^\mu}$\/}.  This can be seen by examining the
transformation changing the complex structure in eq\eq{crot}, and the
definition of the $u$'s in eq\eq{udef}.  The complex structure of
the fiber bundle is thus built into the derivatives.

Now, to describe SDYM using the harmonic formalism, one starts with the
desired solution:  Consider an ordinary four-dimensional vector
potential $A_{\mu i}(x)$ on $R^4$.  The four-dimensional covariant
derivative $\D_{\mu i} \equiv \pa_{\mu i} + A_{\mu i}(x)$ is lifted into
covariant derivatives on the fiber bundle as
\beq
	\D_\mu^\pm \, \equiv \, \pa_\mu^\pm +A_\mu^\pm =
	u^{\pm i} \D_{\mu i} \stop \label{alift}
\eeq
Note that the lifted connection is purely horizontal, \ie{} it has no
components in the direction of the fiber.  Also, $A_\mu^+$ is linear in
$u^{+i}$ and does not depend on $u^{-i}$.  Using the definition of
$D^{++}$ in eq\eq{lie}, and comparing to the expansion of a general
$A_\mu^+$ as in eq\eq{expan}, this condition can be written as $D^{++}
A_\mu^+ = 0$, or
\beq
[ D^{++} , \D_\mu^+ ] = 0 \stop \label{feq}
\eeq
One can return to the four-dimensional gauge field by
projecting $A_\mu^+(x,u)$ to $R^4$.  The projection is performed by
integration:
\beq
A_{{\mu}i} (x) = \int\du \, 2 \, u_i^- A_{\mu}^{+} (x,u) \label{A4} \comma
\eeq
as can be seen using eqs\eq{alift} and (\ref{hint}).
(Note that, since one can find the four-dimensional gauge field from
$A^+$ alone, the corresponding equations for $A_\mu^-$ are
redundant.)

The curvature is lifted in the same way to give:
\beq
\eqalign{
F_{\mu\nu}^{++} & = {\eps}_{\mu\nu} \, u^{+i}u^{+j}F_{(i j)}(x) \cr
F_{\mu\nu}^{--} & = {\eps}_{\mu\nu}\, u^{-i}u^{-j}F_{(i j)}(x) \cr
F_{\mu\nu}^{+-} & = {\eps}_{\mu\nu}\, u^{+i}u^{-j}F_{(i j)}(x)  +
                F_{(\mu\nu)}(x)         \stop
}
\eeq
$F_{\mu\nu}^{++}$, $F_{\mu\nu}^{--}$ and $F_{\mu\nu}^{+-}$
are the $(2,0)$, $(0,2)$ and the $(1,1)$ components
of the curvature, respectively. Note that the $(1,1)$ component
consists of two orthogonal pieces with dimensions one and three, as
in eq\eq{decomp}.

The condition for $F$ to be self-dual, $F_{(i j)}=0$, again
implies that only the irreducible $(1,1)$ piece of $F$ survives.
For the $(2,0)$ component, we have:
\beq
F_{\mu\nu}^{++}=0 \qquad \implies \qquad
[ \D_{\mu}^+  , \D_{\nu}^+ ] = 0    \stop \label{constr}
\eeq
(This equation is the integrability
condition for the equation $\D_{\mu}^+{\phi}=0$.)
As stated in the twistor discussion, the vanishing of this component
for all $u$, corresponding to the vanishing for all complex
structures, is sufficient to show the self-duality of $F$.  This
can be seen by recovering the complete anti-dual part of the four-dimensional
field strength from $F^{++}$:
\beq
F_{i j}(x) = \int\du \, \nfrac32 \, u_i^- u_j^- \eps^{\mu\nu} F_{\mu\nu}^{++}
          \,=\, 0  \stop \label{forg}
\eeq
Thus, eq\eq{constr},
together with the constraint (\ref{feq}) can be regarded as the
equations of motion of the theory \cite{harmsdym}.

\subsect{The action for SDYM in harmonic space}

To get an action for SDYM, it is easiest to first solve eq\eq{constr}.
The general solution is
\beq
\D_{\mu}^+(x,u) = e^{-v(x,u)}{\pa}_{\mu}^+e^{v(x,u)} \stop \label{vdef}
\eeq
With this definition of $\D_\mu^+$, the remaining equation of motion,
eq\eq{feq}, becomes
\beq
\left[ \, D^{++} , e^{-v} \pa_\mu^+ e^{v} \, \right ] = 0
	\quad \iff \quad
     \left[ \, e^v D^{++} e^{-v} , \pa_\mu^+ \, \right ] = 0 \stop
\eeq
Kalitzin and Sokatchev proposed an action for SDYM theory that uses
a lagrange-multiplier field $P^{(-3)\mu}(x,u)$ to enforce this condition
\cite{Action}.  $P^{(-3)\mu}$ has a $U(1)$ charge $-3$, and is in the
adjoint of the gauge group. The complete action is
\beq
S_0 = \int \dxf  \, \du \; \Tr \left( \, P^{(-3)\mu} \, \pa_\mu^+ \,
	( e^v D^{++} e^{-v} )  \, \right)      \stop \label{act}
\eeq
The action has the usual gauge transformation
\beq
\eqalign{
e^{v(x,u)}  &\to e^{v(x,u)} e^{\tau(x)} \cr
\Pmu(x,u) &\to  \Pmu(x,u) \comma \label{vgi}
}
\eeq
with a parameter $\tau$ that is independent of the harmonic
coordinates.  Under this transformation, using eq\eq{vdef}, one sees
that the covariant derivative $\D_\mu^+$ transforms as
\beq
\D_{\mu}^+ \to  e^{-\tau(x)} \D_{\mu}^+  e^{\tau(x)} \comma
\eeq
as expected.  The harmonic-space derivative $D^{++}$ is unaffected. Note that
$\Pmu$ does not transform under the gauge transformation, even though it
appears to be in the adjoint of the group\footnote{One can define the field
$\tilde{P}^{(-3)} = e^{-v} \P e^v$, which does transform in the adjoint of the
group.  We shall need this in the background field quantization of the
theory.}.
The action is also invariant  under the $\Pmu$ gauge transformation
\beq
P^{(-3)\mu} \to P^{(-3)\mu} + {\pa}^{\mu+} \,
          b^{(-4)}(x,u) \stop \label{pgi}
\eeq

As expected, variation with respect to $P^{(-3)\mu}$ yields the SDYM
constraints.  The usual problem with lagrange-multiplier actions is that
the equations of motion of the other fields lead to the lagrange-multiplier
field propagating \cite{lorcov}.  What is unusual in this theory is that the
variation of the action with respect to $v(x,u)$ implies the equation
\beq
\pa_\mu^+ \, P^{(-3)\mu} =0 \stop \label{peq}
\eeq
The $\P$ field is therefore completely decoupled from the Yang-Mills field.
Furthermore, because of the invariance (\ref{pgi}),
Kalitzin and Sokatchev argue that there is no nontrivial $\P$ that satisfies
eq\eq{peq}, so $\P$ does not describe a new degree of freedom.
The action $S_0$ therefore should describe the pure SDYM system.

\subsect{Notes.}

The equations of motion derived from the harmonic space action (\ref{act})
are not simply $F^{(2,0)}=F^{(0,2)}=0$ in a 3-complex dimensional space.  The
constraint that $F^{(2,0)}$ vanishes is implemented via the definition
of $\D_\mu^+$ in eq\eq{vdef}.  Once the
equation of motion eq\eq{feq} fixes the form of the lift to $R^4 \times
S^2$, this implies the vanishing of $F_{(i j)}$.  The structure of the theory
is therefore different from that of ref.~\cite{Witten}.  By using the
harmonic expansions of eq\eq{expan}, one can verify that the
action is well defined and does not suffer from the singularities of
ref.~\cite{Witten}.

In addition to the aforementioned gauge transformations, the action
is invariant under the transformations
\beq
\eqalign{
e^v &\to e^\lambda e^v \cr
\Pmu &\to  e^\lambda \, \Pmu e^{-\lambda} \comma \label{extrasym}
}
\eeq
where $\pa_\nu^+ \lambda(x,u) = 0$.  This implies that $\Box \lambda = 0$,
so {\it this transformation is not a gauge invariance.\/}  The significance
of this transformation will be discussed later.

\section{Quantization of SDYM theory}
\indent

The most important aspect of the action $S_0$ can be seen
without dealing with the details of the theory:  While the action is
nonpolynomial in $v$, it is linear in $\Pmu$.  $\P$, therefore, always
appears in the path integral as $\P / \hbar$, so there are $1-l$
external $\P$'s in any $l$-loop Green function.  Therefore, {\it all S-matrix
elements at tree level have one $\P$ with an arbitrary number of $v$'s, all
one-loop elements have an arbitrary number of $v$'s and no $\P$'s, and there
are no higher loop diagrams.\/}  This statement will hold as long as the
theory is one-loop finite, so that the structure of the
action is not spoiled by counterterms.

Since the results of the loop calculations in this theory involve some
subtleties, we shall perform the quantization in two different ways:
first a straightforward quantization, and then a quantization using a
background field approach.  In the next section, we shall give a simple
argument, modulo issues of the measure of the theory, to explain our
result---that the theory is actually a free theory.  The reader who does
not wish to go through the details of the calculation may be reassured
that few of them are necessary for the understanding of the basic issues.

\subsect{Standard quantization}

Because of the two gauge symmetries (\ref{vgi}) and (\ref{pgi}) of
the action, one needs to gauge fix $S_0$.  Appropriate gauge fixing
conditions for the symmetries are \cite{Action}:
\beq
\eqalign{
&\int\du \; v(x,u) = 0 \cr
&{\pa}_{\mu}^-P^{(-3)\mu}=0 \comma   \label{gf}
}
\eeq
respectively.  The total gauge-fixed action is then
$S=S_0+S_{gf}+S_{gh}$, where
\beq
\eqalign{
S_{gf} &= -\int\dxf \, \du \, \Tr \left({\rho}(x)v(x,u) + \Lam(x,u)
        {\pa}_{\mu}^-P^{(-3)\mu}(x,u) \right) \semi \cr
S_{gh} &= \int\dxf \, \du \, \bar{C}^a(x) \left( {\delta}^{a b} -
    \nfrac12 \, f^{a b c} v^c(x,u) + \frac1{12} \, f^{a c e} f^{e d b} v^c
    v^d + \cdots \right) C^b(x) \cr
       &~- \int\dxf \, \du \, \Tr \left( \nfrac12 \, \chib(x,u) \, \Box
       \chif(x,u) \right) \stop
\label{gf+gh}
}
\eeq
Here $\rho(x)$ and $\Lam(x,u)$ are the Landau-gauge lagrange multipliers,
$C(x)$ and $\chif(x,u)$ are the ghosts and $\bar{C}(x)$ and $\chib(x,u)$
are the anti-ghosts.

The propagators are \cite{Action}\footnote{Here,  ${u_1^+u_2^+}$
denotes ${u_1^{+i} u_{2\,i}^+}$ \etc}:
\beq
\eqalign{
\langle v^a(-p,u_1) ~ P^{(-3)\mu b}(p,u_2) \rangle   & = 2i \, \delta^{a b}
      \, \frac{p_2^{\mu-}}{p^2} \, \frac{u_1^+u_2^-}{u_1^+u_2^+} \cr
\langle P^{(-3)\mu a}(-p,u_1) ~ \Lambda^{(+4)b})(p,u_2) \rangle   & = 2i \,
   \delta^{a b} \, \frac{p_1^{{\mu}+}} {p^2} \, \delta^{(-4,4)}(u_1,u_2) \cr
\langle v^a(-p,u) ~ \rho(p) \rangle   & = -\delta^{a b} \cr
\langle \chi^{(-4)a} (-p,u_1) \bar\chi^{(+4)b}(p,u_2) \rangle  &=
      2\, \frac{\delta^{a b}}{p^2} \delta^{(-4,4)}(u_1,u_2) \cr
\langle C^a(-p) ~ \bar{C}^b(p) \rangle   & = -\delta^{a b}
      \label{props} \stop
}
\eeq
Here, ${\delta}^{(q,-q)}$ and ${u_1^+u_2^-}/{u_1^+u_2^+}$ are harmonic
distributions that are singular when $u_1=u_2$.  They are related by
\beq
D_1^{++}\, \frac{u_1^+u_2^-}{u_1^+u_2^+} = {\delta}^{(2,-2)}(u_1,u_2) \comma
\eeq
and are defined by their series expansions.  For example, the harmonic space
delta-functions are defined as \cite{harm}:
\beq
{\delta}^{(q,-q)}(u_1,u_2) = \sum_{n=0}^{\infty}(-1)^{n+q}\frac{(2n+q+1)!}
{n!(n+q)!}(u_1^+)_{(n+q}(u_1^-)_{n)}(u_2^+)^{(n}(u_2^-)^{n+q)} \stop
\eeq

The theory contains two types of vertices, as can be read from
eqs.~(\ref{act}) and (\ref{gf+gh}): The first type contains
an arbitrary number of $v$ fields coupled to a $\P$.  The second consists of
a $\bar{C}$, a $C$ and an arbitrary number of $v$ fields.
As explained  above, all
physical one-loop diagrams contain only external $v$'s, with something
running around the loop.  Since there are no vertices involving the $\Lam$
or $\rho$ fields,  and since $\chif$ and $\chib$ are free, none of them
appear in diagrams. There are thus only diagrams involving a $C$-ghost loop
and diagrams with a loop of $\langle v\,\P \rangle $ propagators.  Generic
examples of the two types of diagrams are shown in Fig.~1:


Consider first the ghost diagrams:  Note that, since the $\tau$ gauge
fixing did not involve derivatives, the ghost propagator is trivial in
momentum space and the ghost vertex has no momenta.  The integration over
the loop momenta is therefore trivial, and simply gives a factor of $\int
d^{\p1 4}\n2 p / (2\pi)^4 = \delta^4(0)$.  As an example, the 2-point
ghost-loop correction to the effective action is\footnote{$f^{abc}f^{dbc}
= c_v {\delta}^{ad}$.}:
\beq
\delta S_2 = -\frac{c_v}{24} \int \duone \, \dutwo \, \dxf \,
           \Tr \, \left( v(x,u_1) v(x,u_2)\,\right)  \delta^4(0)  \stop
\eeq
The effective action contains infinitely many
such terms, with arbitrary numbers of $v$ fields.  These terms are not very
pretty geometrically, since they are nonlocal in harmonic space.  However,
if one use dimensional regularization, which is the only gauge invariant
regularization available, ${\delta}^4(0)\to 0$ and all these diagrams
vanish.

The second type of diagram vanishes for a similar reason:  In this case, the
$\langle v\,\P \rangle $ propagator does contain a $1/p^2$ factor.  However, in
the $\P v^n$
vertices, $\P$ always appears in the form $\pa_\mu^+ \, \Pmu$ (see
eq\eq{act}).  One therefore only needs the effective propagator
\beq
\langle v^a(-p,u_1) ~~ p_\mu^+ P^{(-3){\mu}b}(p,u_2) \rangle  \, = \,
	-i \, {\delta}^{a b} \; \frac{u_1^+u_2^-}{u_1^+u_2^+} \comma
\eeq
which is again trivial in momentum space.  This is not surprising, since
$p_\mu^+ \Pmu = 0$ by the equation of motion of $\P$ (eq\eq{peq}).  The
remaining part of the vertex (with the $p_\mu^+$ factor removed) also has no
space-time momenta, so the diagrams again all contain a factor of
$\delta^4(0)$.

There is, however, one important difference between the ghost and the $\P v$
diagrams.  Since the $\langle v \P \rangle $ propagator is nontrivial
in harmonic space, the $P v$ diagrams contain complicated, and
singular, harmonic space factors, such as $\delta^{(0,0)}(u,u)$ and
$1/(u^+ u^+)$. If we again use dimensional regularization in spacetime,
and assume that the harmonic-space divergences can be
regulated\footnote{For example, $\zeta$-function regularization
implies that $\delta^{(0,0)}(u,u) = \sum_l (2l+1) \to 1/12$. Alternatively,
one can regularize by restricting the
sum to $l \le L$.  However, it is not clear how to perform these
regularizations consistently for the different harmonic-space
factors.}---so that $0 \times \infty \to 0$---we can conclude that the
one-loop corrections to the effective action vanish.   The theory then
requires no counterterms, so our argument that  higher-loop corrections
do not exist is valid.  We conclude that the tree-level theory is exact.
However, since this conclusion is based on somewhat
delicate arguments, it will be useful to consider the background-field
quantization to support it.

\subsect{Quantum corrections in the background-field formalism}

In the background-field formalism, one splits the fields of the
theory into classical and quantum pieces.  Thus,
\beq
\eqalign{
e^v   & \to e^{v_{cl}} e^{v}  \cr
P^\mu &\to e^{v_{cl}} (P_{cl}^\mu + P^\mu) e^{-v_{cl}}  \stop
}
\eeq
These slightly nontrivial splittings have been chosen to get nice
gauge-transformation properties of the fields.   Substituting these
definitions into the action of eq\eq{act}, one obtains the background-field
action
\beq S_0 = \int\dxf \, \du \; \Tr \left( \, (P^{(-3)\mu} + P_{cl}^{(-3)\mu})
     \left[ \, \nabla_\mu^+ \, , \, (e^{v} D^{++} e^{-v}) \, \right] \, \right)
     \comma  \label{bfact}
\eeq
where
\beq
\nabla_\mu^+ = e^{-v_{cl}} \, \pa_\mu^+ \, e^{v_{cl}} \; \equiv \;
\pa_\mu^+ + A_{\mu \, cl}^+     \comma
\eeq
in analogy to eq\eq{vdef}.

{\it Note that $v_{cl}$ appears in the action
only in the combination $ A_{\mu \, cl}^+(v) $.\/}  This means that the
geometrical meaning of the effective action---and the structure of possible
counterterms---will be much more transparent in the background field approach.
Indeed, $S_0$ is invariant under the classical $\tau(x)$ transformation
\beq
\left(\begin{array}{c}
e^v \\ \nabla_\mu^+ \\ P^{(-3)\mu} \\ P_{cl}^{(-3)\mu}
                           \end{array}\right)
\to e^{-\tau_{cl}(x)}
\left(\begin{array}{c}
e^v \\ \nabla_\mu^+ \\ P^{(-3)\mu} \\ P_{cl}^{(-3)\mu}
                           \end{array}\right)
e^{\tau_{cl}(x)}
\comma  \label{cvgi}
\eeq
under which $A_{\mu\,cl}^+$ transforms as a connection.  $S_0$
is also invariant under the classical $b_{cl}^{(-4)}(x,u)$ transformation
\beq
P_{cl}^{(-3)\mu} \to P_{cl}^{(-3)\mu} + \nabla^{\mu+} \,
          b_{cl}^{(-4)}(x,u) \comma \label{cpgi}
\eeq
where we have used the fact that
\beq
F_{\mu\nu\, cl}^{++} = [ \nabla_\mu^+  , \nabla_\nu^+ ] \, \equiv \, 0
               \stop \label{nof++}
\eeq

The action is also invariant under the quantum gauge transformations:
\beq
\eqalign{
e^{v} & \to e^v e^{\tau (x)} \cr
P^\mu & \to P^{\mu} + {\nabla}_{\mu}^+ \, b(x,u)   \comma
}
\eeq
which need to be gauge-fixed.  Since we want the full effective action
to be invariant under the classical gauge transformations, the gauge
choice must be covariant under background transformations.  The gauge-fixing
conditions (\ref{gf}) therefore become
\beq
\eqalign{
\int\du \; v  & = 0 \cr
{\nabla}_{\mu}^-P^{\mu}  & = 0    \label{bfgf}   \stop
}
\eeq
{\it The second gauge fixing one requires the introduction of the classical
field $A_{\mu \, cl}^-$, which is not in the original theory.\/}   The only
requirement that needs to be satisfied in defining $A_{\mu\,cl}^-$ is that
it  transforms in the required way.  Such a field can be defined but it is
not unique.   For example, two possible definitions are $\nabla_{\mu\,cl}^-
\equiv e^{-v_{cl}} \pa_\mu^- e^{v_{cl}}$, and $A_{\mu\,cl}^- = u^{-i} A_{\mu
i\,cl}(x)$, with $A_{\mu i\,cl}(x)$ defined by projection as in eq\eq{A4}.
It is a check
of the theory that a particular definition of $A_{\mu\,cl}^-$ should not be
needed, since $\nabla_{\mu\,cl}^-$ appears only in the gauge-fixing part of the
lagrangian.  In fact, $A_{\mu\,cl}^-$ should not appear at all in the
effective action.

The gauge-fixing and ghost actions are now
\beq
\eqalign{
S_{gf} &= -\int\dxf \, \du \, \Tr \left({\rho}(x)v(x,u) + {\Lam}(x,u)
        {\nabla}_{\mu}^-P^{(-3)\mu}(x,u) \right) \semi \cr
S_{gh} &= \int\dxf \, \du \, \bar{C}^a(x) \left( {\delta}^{a b} -
    \nfrac12 \, f^{a b c} v^c(x,u) + \nfrac1{12} \, f^{a c e}f^{e d b}v^c
    v^d + \cdots \right) C^b(x) \cr
       &~- \int\dxf \, \du \, \Tr \left( \chib(x,u) \nabla_\mu^- \nabla^{\mu+}
	\chif (x,u) \right)     \stop \label{bgf+gh}
}
\eeq
In order to do one-loop calculations, one needs only the part of the
action quadratic in the quantum fields.  This gives the vertices
shown in Fig.~2.  The propagators are still given by eq\eq{props}. Note
that the $\chif$ ghosts are no longer abelian, and do not decouple.
However, since there are no vertices between $\rho$, $C$ or $\bar{C}$
and the quantum fields,
these fields do not appear in the calculation.



Let us first consider the one-loop two-point functions. The diagrams
for them are depicted in Fig.~3:
Diagram (a), with the loop of $\langle v \P \rangle $ propagators, gives a pure
$A^+ A^+$ contribution to the effective action.  Similarly, diagram
(b), with the $\langle \P \Lam \rangle $ loop, gives a pure $A^- A^-$
contribution.  {\it These terms are exactly canceled by the $A^+ A^+$ and $A^-
A^-$ parts of the ghost diagram (d).}   This means that the effective action
contains only $A^+ A^-$ terms, from the mixed
diagram (c) and the remains of the ghost diagrams.  The resulting 2-point
effective action is
\beq
\delta S_2 \, = \, -\nfrac12 \int \dpf \int \du \,\delta^{00}(u,u) \Tr \left(
	F_{cl\, quad}^{++} (-p,u) \; F_{cl \,quad}^{--} (p,u)
	\int \dlf \frac1{l^2(l+p)^2} \right)  \comma \label{A2piece}
\eeq
where ``quad'' indicates the part of $F_{cl}$ quadratic in $A_{cl}$.
Since $F_{cl}^{++}\equiv 0$, $F_{cl\, quad}^{++}$ contributes only
terms cubic in $A_{cl}$.  The one loop 2-point correction to the
effective action therefore vanishes\footnote{In principle, one again
has to regularize the harmonic divergences.  However, in this case the
divergence multiplies something that vanishes identically.  Also, there
is only one divergence in the background field case, since $1/(u^+u^+)$
terms never appear.  One therefore expects no subtleties in the
regularization.}.

We can now continue to the 3-point functions.  The diagrams are similar to
the 2-point case, and will not be shown explicitly.  As before, the
pure $(A_{cl}^+)^3$ and $(A_{cl}^-)^3$ terms are canceled between the
ghost and nonghost diagrams.  The remaining diagrams give two
contributions.  The first completes the $F_{cl\,quad}$'s in eq\eq{A2piece}
towards the full $F_{cl}$'s.  The second gives terms schematically of the form
\beq
\delta S_3 \, \sim \, \int \du \,\delta^{00}(u,u) \, \Tr \left(
	F_{cl \,quad}^{++} F_{cl \,quad}^{--} \left(
        k^- A_{cl}^+ + k^+ A_{cl}^- \right)  \, \right) \int \dlf
	\frac1{l^2(l-p)^2(l+r)^2}\comma
\eeq
where $k^\pm$ are some combinations of the momenta.  As in the 2-point
case the identity $F_{cl}^{++}\equiv 0$ leads to the vanishing of the 3-point
one-loop effective action.

Note that, as desired, $A_{\mu\,cl}^-$ does not appear
in the quadratic or cubic
parts of the effective action, since they vanished using just the
definition of $A_{\mu\,cl}^+$.
It is not easy to show that this property persists to the $n$-point
function, but it should by the general argument given above.
The only remaining calculation, therefore, is that of the $(A_{cl}^+)^n$
piece of the effective action.  As before, one sees that
this cancels between the  ghost and nonghost diagrams.  The one-loop
effective action therefore vanishes identically.

In fact, this result could have been foreseen using just the background gauge
invariance.  As discussed above, the one-loop action contains only $v$ fields,
with no $\P$ fields, and the $v$ fields have to appear in the form
$A_\mu^+$.  Since the only gauge-invariant function of $A_\mu^+$ is
$F^{++}$, which vanishes identically, the
one-loop effective action must vanish in this gauge.  The explicit calculation
has shown that there are no anomalies to upset this argument.

The background-field calculation has therefore confirmed the result of the
straightforward quantization, and has justified the use of the
dimensional-regularization argument.  We remind the reader that the
vanishing of the one-loop S-matrix means that all higher loop contributions
also
vanish, so the tree-level action is exact.

\section{The S-matrix and the spectrum of the harmonic-space theory}

\indent

We now turn to the classical S-matrix which we have shown to be the full
S-matrix of the theory.  As was seen in the beginning of section 3, the
S-matrix elements will have one external $\P$ field and an arbitrary number of
$v$ fields.  In order to calculate an S-matrix element, one needs to know what
states to put on the external legs.  At this point, one might wonder why the
S-matrix elements have an external $\P$ field, since classical SDYM is
supposed to be described by the $v$ field alone.
(The $\P$ field was introduced simply as a lagrange multiplier.)  Indeed, the
presence of this field as an external state implies the triviality of the S
matrix:


Consider a generic S-matrix element, depicted in Fig.~4.  All the $\Pmu$'s in
the diagram appear in the form $\pa_\mu^+ \Pmu$.  In the internal lines, this
implies that the diagram is local in harmonic space.  However, on the external
lines, this factor of $\pa_\mu^+$ acts on the $\Pmu$ wave function.  Since
the equation of motion of $\Pmu$ is $\pa_\mu^+  P^{(-3)\mu} =0$,
the diagram vanishes.  Therefore, the S matrix is trivial.

The triviality of the S matrix leads one to the conclusion that one
should be able to describe the theory in terms of free fields.  Indeed,
in terms of the field
\beq
V^{++}(v) \equiv e^v D^{++} e^{-v} \comma \label{Vdef}
\eeq
the action of eq\eq{act}
\beq
S_0 = \int \dxf  \, \du \; \Tr \left( \, P^{(-3)\mu} \, \pa_\mu^+ \, V^{++}
  \right) \label{actp}
\eeq
is quadratic.  Therefore, if one could take $\Vp$ as the variable of the
path integral, instead of the field $v$, the theory would obviously be
free.  Note that the transformation $v \leftrightarrow \Vp$ is local in
spacetime.  Also, it is almost one to one.  Different $v$'s that differ
by gauge transformations lead to the same $\Vp$, since $\Vp$ is gauge
invariant.  However, at least perturbatively, this is the only
degeneracy of the mapping.

In fact, as pointed out in ref.~\cite{Zup}, it is in some sense more natural
to take $\Vp$ as the basic field in harmonic-space SDYM.  (Indeed, this
representation has been used in attempts to construct multi-instanton solutions
\cite{multi}.)  That this can be done is seen by performing a
change of frame from the original ``$\tau$ frame'', where covariant
derivatives transform under $\tau(x)$ transformations, to the ``$\lambda$
frame'', where they transform only under the $\lambda(x,u)$
transformations.
In the $\tau$ frame, one has the flat derivatives $D^{++}$ and $D^{--}$, and
the curved derivatives $\D_\mu^\pm$, with $\D_{\mu}^+ = e^{-v} {\pa}_{\mu}^+
e^v$.  The change of frame consists of transforming all derivatives as
$\pa^{(\lambda)} \equiv e^v \pa^{(\tau)} e^{-v}$.  As a result of this,
$\D_\mu^+$ becomes flat in the $\lambda$ frame, while  $D^{++}$ and
$D^{--}$ pick up the connections $\Vp$ and $\Vm$.

$\Vp$ can now be taken as the fundamental field, from which all other fields
can be derived.  The condition that $A_\mu^{+(\tau)}$ be the lift of a
space-time connection, eq\eq{feq}, becomes
\beq
\left[ \, \D^{++} , \D_\mu^+ \, \right ] = 0 \quad \implies \quad
(-) \pa_\mu^+ \, \Vp =0 \comma \label{anal}
\eeq
which is the equation of motion of the action (\ref{actp}).  $\Vm$ is
derived from the constraint
\beq
\left[ \, \D^{++} , \D^{--}\, \right ] = D^0 \quad \implies \quad
     \D^{++} \, V^{--} = D^{--} \, \Vp  \stop
\eeq
Note that this equation has no spacetime derivatives.  Similarly,
$A_\mu^{-(\lambda)}$ is determined from the constraint
\beq
\left[ \, \D^{--} , \D_\mu^+ \, \right ] = \D_\mu^- \quad \implies \quad
     A_\mu^{-(\lambda)} = - \, \pa_\mu^+ \, V^{--}  \stop
\eeq
The field $V^{++}$, satisfying the equation of motion of eq\eq{anal},
thus gives
a free field realization of SDYM.    As an example of this approach,
the one-instanton $\sut$ solutions with scale $\rho$ centered at $x=0$  are
described by \cite{multi}
\beq
V^{++} \u a \d b = x^+\u a x^+ \d b \comma
\eeq
where, as usual, the internal $\sut$ is now identified with $\sutl$ of
spacetime.

\subsect{The spectrum of the harmonic theory}

Since the S matrix of the harmonic theory is trivial, the only
information in the theory is its spectrum.  This is determined by the
solutions to the equations of motion (\ref{anal}) and (\ref{peq}),
which state that $\pa_\mu \, \Pmu=0$ and that $\Vp$ is analytic in
harmonic space ($X$ is analytic if $\pa_\mu^+ \, X =0 $).  The theory
also has the $b^{(-4)}$ gauge invariance of eq\eq{pgi}: $\delta
P^{(-3)\mu} ={\pa}^{\mu+} \, b^{(-4)}$.  The solutions to these
equations should be that $\Pmu$ is trivial, and that $\Vp$ describes
the self-dual configurations.

One should now recall the $\lambda(x,u)$ transformations of eq\eq{extrasym},
under which $\D^{++}$ transforms as a nonabelian connection, and $\Pmu$
transforms covariantly:
\beq
\left(\begin{array}{c} \D^{++} \\ \Pmu \end{array}\right)
\to e^\lambda(x,u) \left(\begin{array}{c} \D^{++} \\ \Pmu
\end{array}\right)  e^{-\lambda(x,u)}
\stop \label{extrasymprime}
\eeq
Since $\lambda(x,u)$ is restricted to be analytic, and is not an
arbitrary function, these transformations
should not be considered as gauge invariances of the action.
Rather, they should be regarded as symmetry transformations linking
different solutions of the theory.  Since the four-dimensional gauge fields are
invariant under these transformations, all configurations related by
$\lambda$ transformations give the same four-dimensional SDYM configuration.
We conclude that the harmonic space action describes an infinite number of
copies of SDYM, and may also have other degrees of freedom.

A detailed study of the spectrum of the harmonic-space theory is
difficult to carry out in Euclidean space, since on $R^4$ the existence
of solutions involves the behaviour of the fields at infinity.  It is
easier to consider the theory in $(2,2)$ Minkowski space, where SDYM
describes one propagating degree of freedom.  In order to do this we
shall not be concerned with global questions, but shall simply regard the
theory as being defined by its harmonic expansion.  (See eq\eq{expan} for the
expansion of a typical field.)  Going to Minkowski
space then means changing the reality properties of the fields, so that
tensors of $\sut$ become tensors of $\sltr$.  The spectrum of the
theory can now be easily analyzed.  One sees that in $(2,2)$ space,
the $\Vp$ field describes three degrees of freedom at the first level of its
expansion ($\Vp \equiv V_{i j} u^{i +} u^{j +} + \cdots$) and one new
degree of freedom at each further level.  One of the
lowest level fields gives the four-dimensional gauge field, while all the
rest can be obtained by $\lambda$ transformations.  One also finds that
the $\Pmu$ field is {\it not}\/ trivial, but describes one free massless
degree of freedom at each level in its harmonic expansion.  We
conclude that {\it the harmonic space theory is a free theory describing
infinitely many copies of SDYM \footnote{Therefore, the theory as such
is not a counterexample to the ``no go theorem'' of ref.~\cite{lorcov}.},
together with extra free particles that are decoupled from the SDYM.
It is not the theory of the $N=2$ string.}

\section{Conclusions}

\indent

The harmonic space description of SDYM has been seen to be closely related to
the twistor construction of Yang-Mills instanton solutions
\cite{Ward,inst,Atiyah}.   In particular, the solutions to the harmonic
equations of motion provide a lift of self-dual configurations on a
four manifold to the projective twistor space.   The theory therefore
has many of the geometrical properties that would be desired in an
action formulation of SDYM.  The fundamental fields of the harmonic
action, $v$ or $\Vp$, are related to the four-dimensional gauge fields
in a nontrivial way.  (See eqs.~(\ref{Vdef}), (\ref{vdef}) and
(\ref{A4}).)  In particular, there are infinitely many $v$ or $\Vp$
fields corresponding to a particular four-dimensional configuration,
since the $\lambda(x,u)$ transformations of eqs.~(\ref{extrasym})
leave the gauge field invariant.  This means that
the action for the theory describes not pure SDYM, but infinitely many
decoupled copies of it.  In addition, in a space with a $(2,2)$ signature,
the action contains an infinite number of free scalar fields.

The complicated redefinition of $A_{\mu i}(x) \to \Vp(x,u)$ leads to a
great simplification to the theory.  Indeed, the action is quadratic in
terms of the fields $\Vp$ and $\Pmu$.  The theory therefore has the
remarkable property of being able to describe the nonlinear interacting
SDYM system as a free theory.  It can do this since, while the theory
is free, the $\lambda(x,u)$ transformations that map equivalent
solutions to each other are nonlinear.  Unfortunately, while the theory
does give a simple description of classical SDYM, its S matrix is
trivial.  It therefore does not describe the $N=2$ string, which
contains a non-trivial 3-point S-matrix element at tree level.

Since the harmonic theory is not relevant to the $N=2$ string, we shall
briefly survey the remaining actions that have been proposed for SDYM.
The first possibility---considered and dismissed in
ref.~\cite{lorcov}---is to simply enforce the self-duality of $F_{\mu
\nu}(A)$ with a lagrange multiplier $\Lambda^{(-)\mu\nu}$.  Thus
\beq
S = \int \dxf \, \Tr \left( \Lambda^{(-)\mu\nu} ( F-\tilde{F} )_{\mu\nu}
\right) \stop \label{naive}
\eeq
Here $\Lambda^{(-)}$ is an anti-self-dual 2-tensor in the adjoint of the
group.  The equation of motion of $\Lambda^{(-)}$ gives SDYM.  However,
the equation of motion of $A_\mu$ shows that $\Lambda^{(-)}$ itself
also propagates.  What is worse is that, unlike the harmonic case,
$\Lambda^{(-)}$ is coupled to the gauge field since it is in the
adjoint of the group.  The theory therefore describes SDYM interacting
with other fields, and it is not appropriate for quantum calculations in
SDYM or the $N=2$ string.

The two remaining actions that have been proposed, both by Nair and
Schiff \cite{NS}, are written in a K\"ahler four-dimensional space
times a line segment parameterized by $t \in [0,1]$.  The boundary at
$t=0$ is taken to be spacetime.  These theories have the disadvantage
that when the four manifold is taken to be $R^4$ one does not have
manifest Lorentz invariance, since a particular complex structure must
be singled out.  The first action is written in terms of a
five-dimensional connection form $A$ and the lagrange multiplier fields
$\Phi$ and $\bar\Phi$, which are $(2,0)$ and $(0,2)$ spacetime forms
cross $d t$, respectively, and are both in the adjoint of the gauge group.
The action is
\beq
S = \int \dxf \, {\rm d} t \, \Tr \left( -\frac{n}{4\pi}
\left ( A {\rm d} A+\nfrac23 A^3 \right ) k +
\Phi F +\bar\Phi F \right) \stop \label{ns1}
\eeq
Here $n$ is an integer, $F$ is the field strength of $A$ and $k$ is the
K\"ahler form on the four manifold.  The lagrange multiplier fields
enforce the condition that $F$ is a $(1,1)$ form, and the equation of
motion of $A_t$ implies that $F \wedge k =0 $, so that $F$ is
self-dual.  The theory therefore gives the equations of motion of
SDYM.  However, the other equations of motion describe the time
components of $F$ in terms of $\Phi$ and $\bar\Phi$, and lead to the
result that $\Phi$ and $\bar\Phi$ satisfy the four-dimensional
equations of motion $\nabla^2 \Phi = \nabla^2 \bar\Phi =0$.  In
\cite{NS} it is argued that, at least on appropriate euclidean four
manifolds, there are no such scalars, so the classical theory describes
pure SDYM.  However, this is certainly not the case in $(2,2)$ space,
where $\Phi$ and $\bar\Phi$ are propagating fields coupled to the gauge
field.  This theory therefore has the same problem as the naive
lagrange-multiplier action of eq\eq{naive}, and also is not suitable for
quantum calculations.

The other action of Nair and Schiff is a K\"ahler version of the
two-dimensional Wess-Zumino-Witten action.  It is written in terms of
$J=g^2=e^\phi$, where $\phi$ is the scalar of the $N=2$ string, so it
contains only the fields of the string theory.  Its equation of
motion is eq\eq{EofM}, which is the equation of motion of Yang
\cite{Yang}, so it is a description of SDYM.  The action is
\beq
S(J) = - \, \frac{n}{4\pi} \int_{{\cal M}^4}
\sqrt{g} g^{\mu\nu} \; \Tr
\left( J^{-1} \pa_\mu J \, J^{-1} \pa_\nu J \right) +
\frac{i n}{12\pi} \int_{{\cal M}^5} \Tr \left ( J^{-1} {\rm d} J \right)^3
\wedge k \stop \label{ns2}
\eeq
Here $J$ is again defined on the five dimensional surface, and is fixed
to some $J_0$ on the boundary $t=1$.  Since this action gives the
classical equation of motion of the string \cite{OVh,prep} (at least up
to the four-point functions), it is the only successful candidate for a
field theory of the heterotic or open $N=2$ string.   It is the analogue
of the Plebanski action for self-dual gravity \cite{Plebanski}, which
gives a scalar field theory for the closed $N=2$ string \cite{OV} in terms
of the K\"ahler potential of the space.

Unfortunately, these theories do not give the correct quantum
amplitudes of the $N=2$ strings.  In the closed string case, this has
been seen in calculations of the partition function \cite{part} and of
the one-loop three-point function \cite{italians}.  Thus, the partition
function of the string is $1/4\pi \int d \tau d \bar\tau /\tau_2^2$,
which is the partition function expected from a scalar in {\it two\/}
rather than four dimensions.  Similarly, the string calculation of the
three-point functions is equal to that of the field theory only if the
loop integrations are carried out in two rather than four
dimensions.  There have been several suggestions proposed for resolving
this issue, but none of them have been successfully implemented.  This
remains a basic problem in the present understanding of the $N=2$
string.

\vskip 1in

{\large \bf \noindent Acknowledgments}

We would like to thank A.B.~Zamolodchikov for a lengthy discussion.
N.M. would like to thank S.~Mandelstam, and Y.O. would like to
thank C.~Vafa and E.~Witten for discussions.

\vskip 1in

\parindent=0pt

{\Large \bf Figure captions.}

\vskip 2mm

{\bf Fig.~1} Generic one-loop Feynman diagrams.

{\bf Fig.~2} Background field vertices.

{\bf Fig.~3} One-loop background field diagrams.

{\bf Fig.~4} A typical S-matrix element.

\end{document}